\documentclass[twocolumn,showpacs,showkeys,amsmath,amssymb,nofootinbib]{revtex4}
\usepackage{graphicx}% Include figure files
\usepackage{dcolumn}% Align table columns on decimal point
\usepackage{bm}% bold math
\usepackage{color}

\newcommand{\beq}{\begin{equation}}
\newcommand{\eeq}{\end{equation}}
\newcommand{\beqn}{\begin{eqnarray}}
\newcommand{\eeqn}{\end{eqnarray}}

\begin{document}
\preprint{APS/123-QED}
\title{Probing Nuclear forces beyond the drip-line using the mirror nuclei $^{16}$N and $^{16}$F} % Force line breaks with \\
\author{I. Stefan$^{1,2,3}$, F. de Oliveira Santos$^{1}$, O. Sorlin$^1$, T. Davinson$^4$, M. Lewitowicz$^1$, G. Dumitru$^2$, J.C. Ang\'{e}lique$^5$, M. Ang\'{e}lique$^1$, E. Berthoumieux$^6$, C. Borcea$^2$, R. Borcea$^2$, A. Buta$^2$, J.M. Daugas$^7$, F. de Grancey$^1$, M. Fadil$^1$, S. Gr\'{e}vy$^1$, J. Kiener$^8$, A. Lefebvre-Schuhl$^8$, M. Lenhardt$^1$, J. Mrazek$^{9}$, F. Negoita$^2$, D. Pantelica$^2$, M.G. Pellegriti$^1$, L. Perrot$^1$, M. Ploszajczak$^1$, O.Roig$^7$, M.G. Saint Laurent$^1$, I. Ray$^1$, M. Stanoiu$^2$, C. Stodel$^1$, V. Tatischeff$^8$ and J.C. Thomas$^1$}

\address{
$^1$ GANIL CEA/DSM-CNRS/IN2P3 Caen, France\\
$^2$ Horia Hulubei National Institute of Physics and Nuclear Engineering P.O. Box MG6 Bucharest-Margurele, Romania \\
$^3$ Institut de Physique Nucl\'eaire Universit\'e Paris-Sud-11-CNRS/IN2P3 91406 Orsay, France.\\
$^4$ Department of Physics and Astronomy University of Edinburgh
Edinburgh EH9 3JZ, United Kingdom \\
$^5$ LPC Caen, ENSICAEN Universit\'e de Caen CNRS/IN2P3 Caen, France\\
$^6$ CEA Saclay Irfu/SPhN F-91191 Gif-sur-Yvette, France \\
$^7$ CEA, DAM, DIF, F-91297 Arpajon, France\\
$^8$ CSNSM CNRS/IN2P3/Universit\'e Paris-Sud B\^at.~104 91405 Orsay Campus, France \\
$^{9}$ Nuclear Physics Institute ASCR CZ-25068 Rez Czech Republic \\
}

\date{\today}% It is always \today, today,

\begin{abstract}
Radioactive beams of $^{14}$O and $^{15}$O were used to populate the resonant states 1/2$^+$, 5/2$^+$  and $0^-,1^-,2^-$  in the unbound $^{15}$F and $^{16}$F nuclei respectively by means of proton elastic scattering reactions in inverse kinematics. Based on their large proton spectroscopic factor values, the resonant states in $^{16}$F can be viewed as a core of $^{14}$O plus a proton in the 2s$_{1/2}$ or 1d$_{5/2}$ shell and a neutron in 1p$_{1/2}$. Experimental energies were used to derive the strength of the 2s$_{1/2}$-1p$_{1/2}$ and 1d$_{5/2}$-1p$_{1/2}$ proton-neutron interactions. It is found that the former changes by 40\% compared with the mirror nucleus $^{16}$N, and the second by 10\%. This apparent symmetry breaking of the nuclear force between mirror nuclei finds explanation in the role of the large coupling to the continuum for the states built on an $\ell=0$ proton configuration.
\end{abstract}

\maketitle

\section{Introduction}

Theoretical description of particle-unbound nuclei in the framework of open quantum systems is a challenge to basic nuclear research \cite{rev2}. In such systems, the coupling to the scattering continuum may lead to the modification of the \textit{effective} interactions \cite{Michel1} and a further reordering of the shells \cite{rev3}. The identification and understanding of the role of specific parts of the nuclear forces  \cite{rev3} in stabilising atomic nuclei and inducing shell evolutions is a central theme of nuclear physics \cite{rev1}. This understanding would bring a better predictive power for exotic nuclei such as those involved in the explosive r-process nucleosynthesis or X-ray bursters where relevant spectroscopic information is not yet available  \cite{Herndl}.
%Exotic nuclei far from the valley of stability are particularly interesting in this respect as with increasing neutron/proton number one eventually reaches the limit of stability of nucleonic matter, the neutron/proton drip line.
A good way to shed light on the effect of the continuum is to compare level schemes of mirror nuclei involving a bound and an unbound nucleus. The asymmetries observed between the mirror nuclei, the so-called Thomas-Ehrman shifts \cite{TES}, can be used to single out the role of nucleon-nucleon interaction.
\par
An ideal case to study the effect of the continuum can be found in the mirror systems: \textit{unbound} $^{16}_{9}$F$_{7}$ and the \textit{bound} $^{16}_{7}$N$_{9}$ nuclei \cite{oga}. In the present work, we propose to determine the energies and widths of the unbound states in $^{15}$F and $^{16}$F using the resonant elastic scattering technique. The new measurements have the advantage to combine excellent energy resolution, high statistics and precise energy calibration that marks a leap in quality and consistency over the precedent results obtained for these nuclei. We shall use these properties to derive the proton-neutron interaction energies in the unbound $^{16}$F nucleus and to determine the role of the continuum in changing effective interactions in mirror nuclei.

\section{Experiment}
The nuclei $^{16}$F and $^{15}$F were studied through the measurement of elastic scattering excitation functions. Elastic scattering reactions at low energies can be described by the Rutherford scattering process, but the excitation function also shows "anomalies", i.e. resonances that are related to states in the compound nucleus. The analysis of these resonances through the R-Matrix formalism \cite{Bert} can be used to determine spectroscopic properties of the states, i.e. excitation energies, widths, and spins, see examples of analysis in Ref. \cite{19Na,18Na}. Four different beams were used in this experiment: the two radioactive beams of $^{15}$O and $^{14}$O for the study of $^{16}$F and $^{15}$F, and the two stable beams $^{14}$N and $^{15}$N for the calibrations.
\par
Radioactive $^{15}$O$^{1+}$ ions were produced at the SPIRAL facility at GANIL through the fragmentation of a 95 AMeV $^{16}$O primary beam impinging on a thick carbon production target. They were post-accelerated by means of the CIME cyclotron up to the energy of 1.2 AMeV, the lowest energy available at this accelerator. Intense (several nAe) stable  $^{15}$N$^{1+}$  and molecular ($^{14}$N$^{16}$O)$^{2+}$  beams came along with the radioactive beam of interest. The selection of one of these species was obtained by using a vertical betatron selection device \cite{Bertrand} located inside the cyclotron, and by choosing the suitable magnetic rigidity of the LISE spectrometer \cite{Ann} after the nuclei have traversed a 38~$\mu$g/cm$^{2}$ carbon stripper foil located at the object focal point of LISE.  It was possible to obtain an $^{15}$O$^{6+}$ beam with an intensity of 1.0(2)x10$^6$~pps and 97(1)~\% purity, or one of two stable beams of $^{15}$N$^{6+}$ or $^{14}$N$^{6+}$ with 10$^8$~pps and 100~\% purity.
\par
The selected ions were sent onto a thick polypropylene (CH$_{2}$)$_{n}$ target in which they were stopped. Some ions underwent proton elastic scattering and the scattered protons were detected promptly to the reaction in a E(300 $\mu$m) silicon detector that covered an angular acceptance of $\pm$1$^\circ$ downstream the target. Resonances in the compound nucleus $^{16}$F were studied through the analysis of the scattered protons spectrum obtained with the $^{15}$O beam. Stable beams of $^{14}$N and $^{15}$N were used for calibration purposes to measure the elastic scattering reactions $^{1}$H($^{14}$N,p)$^{14}$N and $^{1}$H($^{15}$N,p)$^{15}$N in the same experimental conditions. Energy calibrations and resolutions were measured by populating known resonances in the compound nucleus $^{15}$O \cite{15N}: the state J$^{\pi}$=3/2$^+$, E$_R$~=~987(10)~keV, $\Gamma$~=~3.6(7)~keV, and in the compound nucleus $^{16}$O: J$^{\pi}$=0$^-$, E$_R$~=~668(4)~keV, $\Gamma$~=~40(4)~keV and J$^{\pi}$=2$^-$, E$_R$~=~841(4)~keV, $\Gamma$~=~1.34(4)~keV.
\par
In the same manner, radioactive $^{14}$O ions were produced with the intensity of 1.9(1)x10$^5$~pps and post-accelerated to 6~AMeV. The isobaric contamination of the beam was reduced down to 0.0(1)~\%. Scattered protons were detected in $\Delta$E(500~$\mu$m)-E(6~mm cooled SiLi) silicon detectors  that covered an angular acceptance of $\pm$2.16$^\circ$. Resonances in the $^{15}$F compound nucleus were studied through the analysis of the proton spectrum. A pure $^{14}$N$^{6+}$ beam accelerated to 6~AMeV was used for calibrations.
\par
The energy resolution of the measured scattered protons can be determined by the relation: $\sigma_{Lab}~=~\sqrt{\sigma_{det}^{2}+\sigma_{\theta}^{2}+\sigma_{strag}^{2}}$, where $\sigma_{det}$ is the energy resolution of the detector that is $\sigma_{det}$=9~keV (20~keV) in the $^{15}$O ($^{14}$N) setting, $\sigma_{strag}$ is the energy  straggling in the target that is estimated to be lower than ~5~keV from simulations, and $\sigma_{\theta}$ is the energy resolution due to the aperture $d\theta$ of the detector. In inverse kinematics it can be derived that $\sigma_{\theta} =$ tan$(\theta) E d \theta$. Therefore the degradation in energy resolution is minimal when $\theta$~=~0$^{\circ}$. For this reason, and for maximizing the ratio between the nuclear and the Coulomb contribution of the differential cross-section, the scattered protons were measured at forward angles. An energy resolution of  $\sigma_{Lab}$ = 10~keV was measured in the case of $^{16}$F, which leads to  $\sigma_{CM} \simeq$ 3~keV in the center of mass, and  $\sigma_{Lab}$ = 22~keV or $\sigma_{CM} \simeq$ 7~keV in the case of $^{15}$F.

\section{Results}
\begin{figure*}
\resizebox{0.8\textwidth}{!}{%
\includegraphics{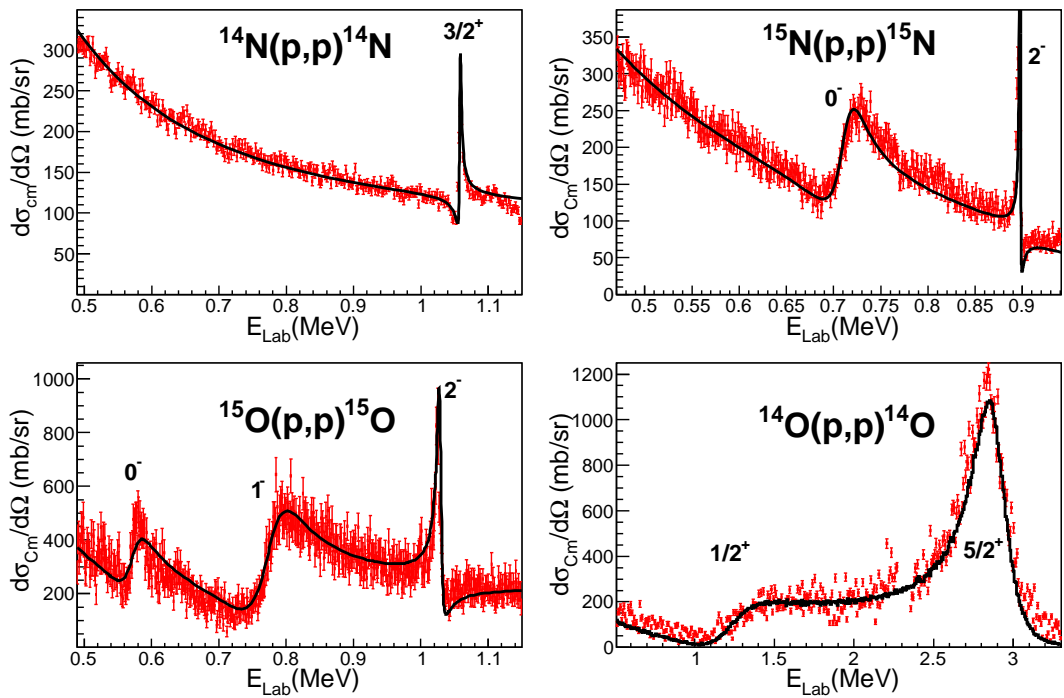}% Here is how to import EPS art
} \caption{\label{spectra} Measured excitation functions of proton resonant elastic scattering on the $^{14}$N, $^{15}$N, $^{15}$O and $^{14}$O nuclei. The differential cross sections measured at 180$^{\circ}$ in the CM are shown as a function of the laboratory energy E$_{Lab}$. Several resonances in the compound nuclei $^{15}$O, $^{16}$O, $^{16}$F and $^{15}$F can be observed. The lines are results of the R-matrix calculations using the parameters from Ref. \cite{15N} for the $^{15}$O and $^{16}$O nuclei, from Table \ref{Table2} for the $^{16}$F nucleus, and from this work for the two resonances observed in $^{15}$F.}
\end{figure*}

The measured scattered protons spectra of $^{1}$H($^{15}$O,p)$^{15}$O and $^{1}$H($^{14}$O,p)$^{14}$O were transformed into the center of mass excitation functions by taking into account the energy losses, the energy and angular straggling, the intrinsic energy resolution and the angular acceptance of the Si detectors. This procedure was successfully tested using the $^{1}$H($^{14}$N,p)$^{14}$N and $^{1}$H($^{15}$N,p)$^{15}$N reactions (see the upper part of Fig. \ref{spectra}). Reactions with the carbon nuclei of the target, although very weak, were subtracted in the case of $^{15}$F using data from Ref. \cite{Lee}, and neglected in the other systems. The incident energies in CM are 1.15~MeV and 5.6~MeV respectively. As the lowest excited states in $^{15}$O and $^{14}$O lie above 5.1 MeV, inelastic scattering do not contribute to the reaction with $^{14}$O, and is expected to be negligible in the case of $^{15}$O.
\par
The spectra corresponding to $^{16}$F and $^{15}$F are shown in the lower part of Fig.\ref{spectra}. They were fitted using the R-matrix formalism with the code Anarki \cite{Bert}. The shape and height of the peaks are used to derive the energy, J$^\pi$ and width of the resonances. The uniqueness of the solution was controlled carefully. Two broad resonances are found at S$_p$=-1.31(1)~MeV, $\Gamma_R$=853(146)~keV, J$^{\pi}$=1/2$^+$ and E$_R$=2.78(1)~MeV, $\Gamma_R$=311(10) keV, J$^{\pi}$=5/2$^+$ in $^{15}$F. Energies and widths of these resonances are consistent with previous results having larger uncertainties \cite{15F}. Three resonances corresponding to the 0$^-$, 1$^-$ and 2$^-$ states in $^{16}$F were identified. Their energy and width are given in Table \ref{Table2}. A proton separation energy S$_{p}$~=~-534~$\pm$~5~keV was obtained, in agreement with the value S$_{p}$ = -536 $\pm$ 8 keV recommended both in the compilation \cite{15N} and the recent measurement S$_{p}$ = -535 keV \cite{Lee}. Conversely, the observed width of the 1$^-$ resonance, 70 $\pm$ 5 keV, differs significantly with the recommended value of $<$ 40 keV or with the recent experimental value of 103 $\pm$ 12 keV \cite{Lee}, and is in good agreement with the most recent measurement of 87 $\pm$ 16 keV \cite{Fujita}. New recommended weighted mean values are proposed in Table \ref{Table2} for the three resonances, while the properties of the J$^\pi$=3$^-$ state are taken from Ref. \cite{Lee}. The measured widths $\Gamma^{exp}_p$ of resonances are related to the spectroscopic factors C$^{2}$S through the relation $\Gamma_{exp}=C^{2}S~\Gamma_{sp}$, where $\Gamma_{sp}$ are the calculated single-particle widths \cite{Lee}. As shown in Table \ref{Table2}, the measured spectroscopic factors of the low-lying states in $^{16}$F are all close to 1. It is also the case in the mirror nucleus $^{16}$N \cite{Guo,Bardayan}.
\par
A recent publication \cite{Wu} reported very precise predictions of the $^{16}$F low-lying states widths using charge symmetry of strong interaction and Woods-Saxon potentials. A 30~\% disagreement is observed between these predictions and our very accurate experimental results. It suggests that a larger value of the radius should be used in the potential. As shown in Fig. 6 of Ref. \cite{Wu}, a value of 3.6 fm gives a much better agreement.

\begin{table*}
\caption{\label{Table2}Measured energies, widths and deduced spectroscopic factors C$^2$S for the low-lying
states in $^{16}$F. The new recommended separation energy is S$_{p}$ = -535 $\pm$ 5 keV. The spectroscopic factors are calculated using the method proposed in Ref \cite{Lee}.}
\begin{ruledtabular}
\begin{tabular}{|ccc|cc|c|cc|ccc|}
\multicolumn{3}{|c|}{Compilation ref \cite{15N}} &  \multicolumn{2}{c|}{Ref \cite{Lee}}&  \multicolumn{1}{c|}{Ref \cite{Fujita}} &  \multicolumn{2}{c|}{This work}&\multicolumn{3}{c|}{New Recommended}\\
\hline
E$_{x}$ (keV) &J$^{\pi}$&$\Gamma_{p}$ (keV)&E$_{x}$ (keV)& $\Gamma_{p}$ (keV) & $\Gamma_{p}$ (keV) & E$_{x} (keV)$&$\Gamma_{p}$ (keV)&E$_{x} (keV)$&$\Gamma_{p}$ (keV)& C$^2$S\\
\hline
0&0$^{-}$&40 $\pm$ 20 &0  &22.8 $\pm$ 14.4 & 18 $\pm$ 16&0 &  25 $\pm$ 5&0&25.6 $\pm$ 4.6&1.1(2)\\
193 $\pm$ 6&1$^{-}$ &$<$ 40 &187 $\pm$ 18&103 $\pm$ 12& 87 $\pm$ 16&198 $\pm$ 10 &  70 $\pm$ 5&194 $\pm$ 5&  76 $\pm$ 5 &0.91(8)\\
424 $\pm$ 5&2$^{-}$ & 40 $\pm$ 30 &416 $\pm$ 20& 4.0 $\pm$ 2.5&16 $\pm$ 16 &425 $\pm$ 2&  6 $\pm$ 3&424.8 $\pm$ 1.9&5.0 $\pm$ 2.0&1.2(5)\\
721 $\pm$ 4 &3$^{-}$& $<$ 15& 722 $\pm$ 16&15.1 $\pm$ 6.7 &12 $\pm$ 16 & & & 721 $\pm$ 4&15.1 $\pm$ 6.7 &1.0(5)\\
\end{tabular}
\end{ruledtabular}
\end{table*}

\section{Interpretation}
\begin{figure*}
\resizebox{0.6\textwidth}{!}{%
\includegraphics{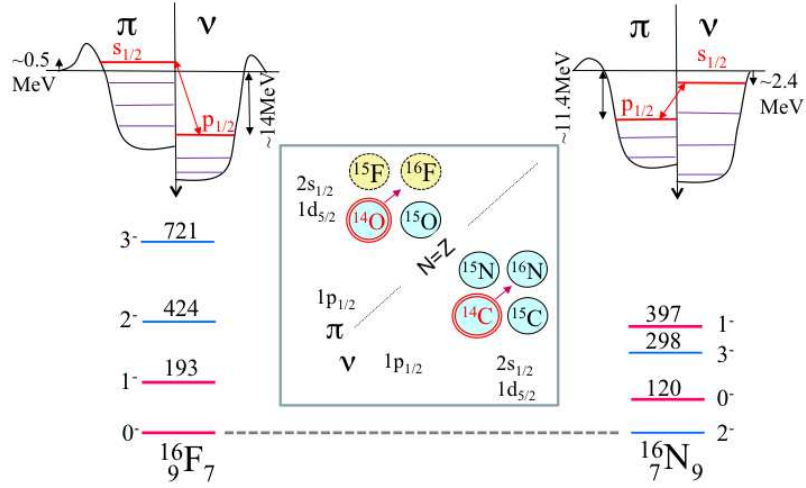}%
}
\caption{Schematic view of the valence proton ($\pi$) and neutron ($\nu$) orbits involved in the mirror systems $^{16}_{9}$F$_{7}$ and $^{16}_{7}$N$_{9}$ on top of the $^{14}_{8}$O$_{6}$ and $^{14}_{6}$C$_{8}$ core nuclei. The major difference between these two systems is that the proton 2s$_{1/2}$ or  1d$_{5/2}$ (not shown) orbits  are unbound in  $^{16}_{9}$F$_{7}$. The level schemes of the two mirror nuclei differ significantly, as shown in the bottom part of the figure.}
\label{mirror}
\end{figure*}

The comparison of the two level schemes of the mirror nuclei $^{16}$N and $^{16}$F is shown in Fig. \ref{mirror}. Large differences can be observed: the ground state of $^{16}$F has 0$^-$ while that of $^{16}$N has 2$^-$, and the two 0$^-$ and 1$^-$ states are down shifted in energy relatively to the other states by more than 500 keV.

\par
In these odd-odd nuclei, a good approach is to use a core plus two nucleons model. The first excited states of the $^{16}_{7}$N$_{9}$ nucleus can be well described using a single-particle description with a closed core of $^{14}$C plus a deeply bound proton in the 1p$_{1/2}$ orbital  (S$_p(^{15}$N)=+10.2~MeV) plus a neutron in the  2s$_{1/2}$ orbital (S$_n(^{15}$C)=+1.22 MeV), leading to J$^\pi$=$0^-,1^-$ states, or plus a neutron in the 1d$_{5/2}$ orbital  (S$_n(^{15}$C*)=+0.48~MeV), leading to J$^\pi$=$2^-,3^-$ states (see Fig. \ref{mirror}). In the same way, (0,1)$^-$ and (2,3)$^-$ states in $^{16}$F can be described as a $^{14}$O core plus a neutron in the 1p$_{1/2}$ orbital  (S$_n(^{15}$0)=+13.22~MeV) plus a proton in the  2s$_{1/2}$ orbital (S$_p(^{15}$F)=-1.31 MeV) or plus a proton in the 1d$_{5/2}$ orbital (S$_p(^{15}$F*)=-2.78~MeV). This simplified single particle view is justified as the spectroscopic factor values of the systems ($^{15}$N = $^{14}$C + p), ($^{15}$F = $^{14}$O + p) and ($^{15}$C=$^{14}$C + n) are close to unity \cite{15N,15C,15F}. In this framework, the experimental neutron-proton (n-p) \textit{effective} interactions elements in $^{16}$N, labeled Int$^{exp}_{^{16}N}$ (J), can be  extracted from the experimental binding energies (BE) as in Ref. \cite{Lepailleur}:
\[
\mathrm{Int^{exp}_{^{16}N}(J)}= \mathrm{BE(^{16}N)_J} - \mathrm{BE(^{16}N_{free})}.
\]
%+ \mathrm{E_{c}(^{16}N)} - \mathrm{E_{c}(^{15}N)}
In this expression BE($^{16}$N$_{free}$) corresponds to the binding energy of the $^{14}$C+1p+1n system without residual interaction between the proton and the neutron. In the case of the J$^\pi$=$0^-$ and $1^-$ states:
\[
\mathrm{BE(^{16}N_{free})} = \mathrm{BE(^{14}C)}_{0^+} + \mathrm{BE(\pi 1p_{1/2})} + \mathrm{BE(\nu 2s_{1/2})}
\]
where
\[
\mathrm{BE(\pi 1p_{1/2})}= \mathrm{BE(^{15}N)}_{1/2^-} - \mathrm{BE(^{14}C)}_{0^+}
\]
\[
\mathrm{BE(\nu 2s_{1/2})}= \mathrm{BE(^{15}C)}_{1/2^+} - \mathrm{BE(^{14}C)}_{0^+}
\]
Combining these equations, we obtain:
\[
\mathrm{BE(^{16}N_{free})} = \mathrm{BE(^{15}C)}_{1/2^+}+\mathrm{BE(^{15}N)}_{1/2^-}-\mathrm{BE(^{14}C)}_{0^+}
\]
The same method has been also applied to obtain the J$^\pi$=$2^-,3^-$ states originating from the n-p coupling $\pi$1p$_{1/2}  \otimes \nu$1d$_{5/2}$. The obtained experimental n-p interaction energies Int$^{exp}_{^{16}N}$ and Int$^{exp}_{^{16}F}$ are given in Table \ref{coupling}.
\begin{table}
\caption{\label{coupling} Experimental proton-neutron interaction energies, Int$^{exp}_{^{16}N}$ and Int$^{exp}_{^{16}F}$, derived for the $J^\pi=0^-,1^-$ [1p$_{1/2} \otimes$ 2s$_{1/2}$] and $J^\pi=2^-,3^-$ [1p$_{1/2} \otimes$ 1d$_{5/2}$] states in $^{16}$N and $^{16}$F. Calculated interaction energies Int$^{over}_{^{16}F}$ and Int$^{over+C}_{^{16}F}$ are  based on the interactions derived for $^{16}$N to which the effects of the change in wave functions overlaps (over) between mirror nuclei and the change in Coulomb (over+C) energy have been added (see text for details).}
\begin{ruledtabular}
\begin{tabular}{|c|c|cc|c|}
State (J) & Int$^{exp}_{^{16}N}$ &  Int$^{over}_{^{16}F}$& Int$^{over+C}_{^{16}F}$  & Int$^{exp}_{^{16}F}$  \\
\hline
 0$^-$& -1.151 & -0.943 & -0.775 &  -0.775 \\
 1$^-$ &  -0.874& -0.716 & -0.581 & -0.577\\
  \hline
2$^-$ & -2.011& -2.031  &-1.842 &  -1.829\\
3$^-$&-1.713&-1.730 &-1.574  &  -1.523\\
\end{tabular}
\end{ruledtabular}
\end{table}
%Interaction energies Int$^{exp}_{^{16}F}$ (J) corresponding to the J$^\pi$=0$^-$,1$^-$ (2s$_{1/2} \otimes$1p$_{1/2}$) and J$^\pi$=2$^-$,3$^-$ states (1d$_{5/2} \otimes$1p$_{1/2}$) are shown in Table \ref{coupling}.
While the effective interactions of the J$^\pi$=0$^-$,1$^-$ states differ by as much as 40\% in the mirror systems, the  interaction energies of the  J$^\pi$=2$^-$,3$^-$ states differ only by 10\%, despite the fact that the proton 1d$_{5/2}$ orbit is less bound than the 2s$_{1/2}$ one by about 1.5 MeV.
\begin{figure}
\resizebox{0.47\textwidth}{!}{%
\includegraphics{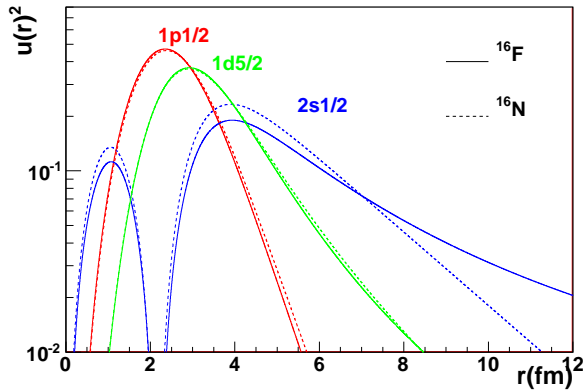}%
} \caption{(colors online) Calculated 1p$_{1/2}$, 1d$_{5/2}$ and 2s$_{1/2}$ single particle radial wave functions u$_n$(r) and u$_p$(r) (see text) for $^{16}$N (dashed lines) and $^{16}$F (full lines). For the scattering states, the unbound proton wave function in the region between r=0 and r=16 fm is renormalized to 1 in order to have the proton inside the nucleus before it decays. The 2s$_{1/2}$ wave function of the unbound proton in $^{16}$F is more spatially spread than it is for the bound neutron in $^{16}$N, contrary to the 1d$_{5/2}$ unbound proton wave function that is retained inside the nucleus by the centrifugal barrier. }
\label{wf}
\end{figure}

\par
Hereafter, we show that these changes of n-p effective interactions can be explained by the effect of the coupling with continuum which leads both to a change in the spatial overlap of the neutron and proton wave functions and in the Coulomb electrostatic energy. Since the n-p effective interaction energies (0.5-2 MeV) are much smaller than the mean nuclear potential energy ($\approx$ 50 MeV), we use the approximation that proton and neutron radial wave functions $u_p(r,J)$ and $u_n(r,J)$ can be calculated by neglecting the n-p interaction, and by solving the Schr\"{o}dinger equation in Woods-Saxon nuclear potentials whose depths have been adjusted to reproduce the observed neutron or proton binding energies for the states in $^{16}$N and $^{16}$F. Results as shown Fig. \ref{wf}. Wave functions are quasi-identical between $^{16}$F (full lines) and $^{16}$N (dashed lines) except for the 2$s_{1/2}$ wave functions. Using a schematic zero-range \emph{v}$_J^{pn}$~=~a$_J$~$\delta$(r$_p$-r$_n$) interaction, the n-p interaction Int$^{over}$(J) is \cite{Heyde}:
\begin{equation}\label{heyde}
Int^{over}(J)= \frac{a_J}{4\pi} \int_0^\infty \frac{1}{r^2}[u_p(r,J)u_n(r,J)]^2dr
\end{equation}
where a$_J$ contains the strength of the n-p nuclear interaction. As the zero-range delta function is only a crude approximation of the nuclear force, the a$_J$ coefficients have been adjusted to equate the calculated and experimental (Int$^{exp}_{^{16}N}$(J)) interaction energies in $^{16}$N. By virtue of the charge symmetry of nuclear forces, the \textit{same} a$_J$ coefficients should be used to calculate the interaction energies Int$^{over}_{^{16}F}$(J) in $^{16}$F. Comparison between experimental and calculated interaction energies given in Table \ref{coupling} deserve several important remarks.
Firstly, the Int$^{over}_{^{16}F}$(2,3) values are similar to those in the mirror system Int$^{exp}_{^{16}N}$(2,3). This can be explained by the fact that the unbound proton and bound neutron 1$d_{5/2}$ wave functions are similar in the mirror systems (see Fig. \ref{wf}). This is due to the high $\ell $=2 centrifugal barrier of $\approx$~3~MeV and Coulomb barrier of $\approx$~3~MeV that prevent the unbound proton in $^{16}$F to couple strongly with the continuum. This is also confirmed with the very narrow measured widths of 5~keV and 15~keV of the 2$^-$ and 3$^-$ states. Secondly, the calculated Int$^{over}_{^{16}F}$(0,1) values, built on the 2$s_{1/2}$ $\otimes$ 1$p_{1/2}$ coupling, are intermediate between the experimental values of $^{16}$N and $^{16}$F. As $\ell$=0 protons (from 2$s_{1/2}$) in $^{16}$F do not encounter any centrifugal barrier, their radial wave function is more extended, as shown in Fig. \ref{wf}. Consequently the overlap between the wave functions of the 2$s_{1/2}$ unbound proton and the deeply bound 1p$_{1/2}$ neutron is reduced. Thirdly, as the Int$^{over}_{^{16}F}$(J) values differ from the experimental Int$^{exp}_{^{16}F}$(J) values, an additional effect is required to understand the differences.

\par
While the change in Coulomb energy between two isotopes can be usually neglected for bound states, the apparent radial extension of the wave function of an unbound state can be larger than that of a bound state, leading to a change in Coulomb electrostatic energy E$_c$(J). In the case of $^{16}$F, the interaction energy should therefore be rewritten for each state J. As for the J=0,1 states it writes:
\[
\mathrm{Int^{over+C}_{^{16}F}(J)}= \mathrm{Int^{over}_{^{16}F}(J)}+ \mathrm{E_{c}(^{16}F(J))}- \mathrm{E_{c}(^{15}F)_{1/2^+}}
\]
Coulomb energies are determined by using the following relation between the charge distribution $\rho(r)$ of the core $^{14}$O nucleus with $Z_{core}$=8 and a single proton in the 2$s_{1/2}$ or 1$d_{5/2}$ orbit having radial wave functions $u_p(r,J)$:
\begin{equation}\label{elec}
    E_c(J)=\frac{Z_{core}}{4\pi\epsilon_0}\int_0^\infty \frac{\rho(r)u_p(r,J)^2}{r}dr
\end{equation}
To give an example, it is found that $E_{c}(^{16}F(0^-))-E_{c}(^{15}F)_{1/2^+}=168~keV$. By applying this Coulomb energy correction as well as the one due to the change in wave functions overlaps between mirror nuclei, it is found that the calculated Int$^{over+C}_{^{16}F}$(J) and experimental Int$^{exp}_{^{16}F}$(J) interaction energies are very similar, as shown in Table \ref{coupling}. This suggests that differences in the proton-neutron interaction energies between the two mirror nuclei are very well accounted for by these two combined effects, the amplitude of which sensitively depends on the energy \emph{and} angular momentum of the states under study.
\par
Our results agree very well with those obtained by Ogawa et al. \cite{oga} where the mirror system $^{16}$F-$^{16}$N was studied using different model which is based on one particle plus one hole on top of the $^{16}$O inert core, the residual interaction being calculated with the M3Y interaction and single-particle wave functions obtained under the Woods-Saxon plus Coulomb potential. Compared to this work, our model is even simpler and explains perfectly the observed differences between the two mirror nuclei. This good agreement and the simplicity of our model makes this system a particularly interesting textbook case for understanding the effect  coupling to continuum on effective nuclear forces and subsequent shell reordering.
\par
As a warning, we would like to add that in our simplified approach, the effects of nuclear correlations to derive the effective nuclear forces in the A=16 mirror nuclei have been neglected. We consider that this assumption is reasonable due to the fact that correlations should be at the first order similar in the two mirror nuclei. We expect that the implementation of nuclear correlations will slightly change the intensity of the derived effective interactions, but this change would be similar in the two mirror nuclei. The role of correlations certainly deserves a dedicated treatment that goes beyond the scope of the present work.

\section{Conclusion}
The 1/2$^+$, 5/2$^+$  and $0^-,1^-,2^-$ resonant states were studied in the $^{15}$F and $^{16}$F nuclei, respectively, with unprecedented energy accuracy and resolution by means of proton elastic scattering reactions in inverse kinematics. Experimental energies were used to deduce that the effective proton-neutron interactions between the mirror nuclei differ by as much as 40\% for the J$^\pi$=0$^-$,1$^-$ states and by only 10\% for the J$^\pi$=2$^-$,3$^-$ states, although the latter states lie at higher energy in the continuum. We demonstrate that these features are well explained by the effect of the presence of the proton wave functions in the continuum that reduces proton and neutron radial overlaps and induces significant changes in Coulomb energy. This mirror system is ideal to test models in which the role of the continuum is or will be implemented, such as in {\em ab initio} no-core shell model \cite{rev4}, coupled cluster approach \cite{CC1}, or shell model approaches \cite{rev2,Signoracci}. A correct treatment of continuum is required  for predicting the location of drip-lines, modeling weakly bound nuclei such as halo nuclei and predicting resonant states of astrophysical importance for X-ray bursters and for the r-process nucleosynthesis.

\acknowledgments {\small
We thank the GANIL crew for delivering the radioactive beams. We thank B. Bastin, C. Bertulani, P. Descouvemont, M. Dufour, F. Nowacky and P. Van Isacker for fruitful discussions. This work has been supported by the European Community FP6 - Structuring the ERA - Integrated Infrastructure Initiative- contract EURONS n° RII3-CT-2004-506065, the IN2P3-IFIN-HH No. 03-33, the IN2P3-ASCR LEA NuAG and partially through FUSTIPEN (French-U.S. Theory Institute for Physics with Exotic Nuclei) under DOE grant number DE-FG02-10ER41700.}

%%\newpage %Just because of unusual number of tables stacked at end
%%\bibliography{apssamp}% Produces the bibliography via BibTeX.

\end{document}